\documentclass[10pt]{article}

\usepackage{amsmath}
\usepackage{amsfonts}
\usepackage{amssymb}
\usepackage{bbm}
\usepackage{color}
\usepackage{graphicx}
\usepackage{pstcol}
\usepackage{amscd}

\begin{document}
\title{Testing the Dirac equation\footnote{To appear in: C. L\"ammerzahl, C.W.F. Everitt, and F.W. Hehl (eds.): {\it Gyros, Clocks, Interferometers...: Testing Relativistic Gravity in Space}, Lecture Notes in Physics {\bf 562}, Springer 2001.}}

\author{Claus L\"ammerzahl$^1$
and Christian J. Bord\'e$^2$ \\ 
\\
$^1$Fakult\"at f\"ur Physik, Universit\"at Konstanz, 78457 Konstanz, Germany
\\
$^2$Laboratoire de Physique des Lasers, Institut Galil\'ee \\ 
Universit\'e Paris 13, 93430 Villetaneuse, France
}

\maketitle

\begin{abstract}
The dynamical equations which are basic for the description of the dynamics of quantum felds in arbitrary space--time geometries, can be derived from the requirements of a unique deterministic evolution of the quantum fields, the superposition principle, a finite propagation speed, and probability conservation. 
We suggest and describe observations and experiments which are able to test the unique deterministic evolution and analyze given experimental data from which restrictions of anomalous terms violating this basic principle can be concluded. 
One important point is, that such anomalous terms are predicted from loop gravity as well as from string theories. 
Most accurate data can be obtained from future astrophysical observations. 
Also, laboratory tests like spectroscopy give constraints on the anomalous terms. 
\end{abstract}

\section{Introduction}\label{ArticleLaemmerzahlBorde}\index{Dirac equation}

Experimental Quantum Gravity, the experimental search for deviations from Einstein's General Relativity, which includes also Special Relativity, has been an very active area in physics since a few years. 
All unifying theories or quantum gravity theories predict small modifications from General Relativity, for example deviations from the $1/r$--potential, and violation of the equivalence principle, see e.g.\ \cite{FischbachTalmadge99}, violation of Lorentz--invariance and violations of the universality of the gravitational red shift. 
Violations of these basic principles underlying General Relativity go together with a modification of the equations of motion for test matter in the gravitational field. 

Questions about the structure of the field equations for quantum objects came
up very recently in the context of \index{quantum gravity} quantum gravity: From loop gravity
\cite{BL_AlfraoMoralesTecotlUrrutia00} as well as from string theory
\cite{BL_Ellisetal99} there are predictions about quantum gravity induced
modifications of the field equations governing the motion of
spin--$\frac{1}{2}$--particles.  Up to now the predictions consist in
directional derivatives and higher--order spatial derivatives added to the
usual Dirac equation.  However, more general modifications can also be
expected.  For example, the prediction of second order spatial derivatives,
which is worked out in a distinguished frame of reference, makes it reasonable
to expect also second order time derivatives.  Therefore, it is important to
study experimental consequences of general non--standard modifications of the
Dirac equation.  In order to have a guiding principle at hand which tells us
something about the meaning and the physical consequences of certain
modifications of the usual Dirac equation, we will study these modifications
in the frame of a constructive axiomatic derivation of the Dirac equation.

In that approach it is possible to derive the Dirac equation from a very few fundamental {\it operational} principles which can be tested directly in experiments. 
The derivation of the Dirac equation can be divided into two parts: The first part consists in the derivation of a system of linear hyperbolic field equations, which we call a generalized Dirac equation, from four fundamental principles which the quantum field has to obey, namely (i) the unique deterministic \index{deterministic evolution} evolution, (ii) the \index{superposition principle} superposition principle, (iii) finite propagation speed, and (iv) conservation of probability: 
\begin{equation*}
\left.\begin{matrix} 
\hbox{unique deterministic evolution} \\ \hbox{superposition principle}\hfill  \\ \hbox{finite propagation speed}\hfill \\ \hbox{conservation law}\hfill \end{matrix} 
\right\} 
\Rightarrow 
\left\{\begin{matrix}
\hbox{generalized Dirac--equation} \\ 
0 = i \gamma^a(x) \partial_a \varphi(x) - M(x)\varphi(x) \end{matrix}
\right.
\end{equation*}
Here $\gamma^a$ ($a = 0, \ldots, 3$) are some matrices which are not assumed to fulfill a Clifford algebra. Also $M$ is a matrix. 

The usual Dirac equation where the matrices $\gamma^a$ fulfill a Clifford algebra and where $M$ is proportional to the unit matrix, can be derived from the additional assumptions 
(v) uniquenes of the null cones, (vi) two helicity states only, and (vii) uniqueness of the mass shell, compare \cite{AL93,LaemmerzahlBleyer99}. 
From these demands we arrive at a Dirac equation in Riemann--Cartan space--time, where the coupling to torsion consists in the axial part only. 
All these principles are operational since they can be proven directly by experiments. 

Some basic features of quantum theory mentioned above have been questioned previously and subsequently been tested or estimated from some existing data on atoms, for example. 
One of these features is the linearity of quantum theory which is basic in our understanding of all quantum phenomena. 
A generalized quantum field equation including a non--linear term has been introduced and discussed by e.g.~Shimony \cite{Shimony79}, and Weinberg \cite{Weinberg89}, and references therein. 
Shimony himself proposed a neutron interferometry experiment which subsequently was performed by Shull et al \cite{Shulletal80} giving a strong restriction on the strength of a hypothetical nonlinear term in the Schr\"odinger equation. 
Also spectral data of the hydrogen atom have been used for this purpose. 
Another features which has been discussed was the conservation of probability \cite{Ellisetal84}. 
The second part of assumptions (v) to (vii) manifests itself in a breaking of
local Lorentz-- and local position invariance which can be tested by
Hughes--Drever, red--shift and atomic interferometry experiments
\cite{LB_Laemmerzahl98}. 

In this paper we want to question the first of the four basic principles (i) to (iv) underlying our basic understanding of quantum theory, namely the unique deterministic evolution. 
Here, unique deterministic evolution means that if a quantum state $\psi(x)$ is prepared at a time $t_0$, then the state is uniquely determined for times $t > t_0$. 
This implies that the evolution of quantum states is described by an evolution equation which contains a first time derivative only, that is, $\frac{d}{dt} \psi = A \psi$, where $A$ is some operator.  

In order to test this principle, we propose a generalization of the usual Dirac equation by adding a second time derivative which violates this principle. 
This modified Dirac equation is used to calculate modifications of the propagation of spin--$\frac{1}{2}$--particles as well as the corresponding modifications of the atomic spectrum (The hydrogen spectrum can also be used as justification of the four--dimensionality of space--time on atomic scales \cite{BurgbacherLaemmerzahlMacias99}). 
Up to now, all the experimental results are well explained using the standard theory, that is, the usual Dirac equation together with quantum corrections. 
If everything is well explained using the standard theory, then the modifications of the results due to the modifications in the Dirac equation, can be only smaller than the experimental error. 
Therefore, all the modifications can be restricted by comparing the calculated effects with the accuracy for the various measured effects. -- However, future observation of neutrinos and high energy photons from gamma ray bursts (GRB) may be capable to distinguish between the various models of the Dirac equation. Nevertheless, one has to keep in mind, that on cosmological distances the parameters may depend on the position and thus the effect we are looking for may be cancelled during the propagation over long distances. 
Therefore, these observations have to be complemented by laboratory experiments like spectroscopy.  

\section{The model: A modification of the Dirac equation}

The unique deterministic evolution implies that the evolution equation for the quantum field is of first order in the time derivative. 
This means especially, that the evolution equation is an equation without \index{memory}memory. 
In terms of a system of partial differential equations this means that this system should be of first order in time as it is the case for the Dirac equation $i \hbar \partial_t \psi = - i \hbar c \mbox{\boldmath$\alpha$} \cdot \mbox{\boldmath$\nabla$} \psi + \beta m c^2 \psi$. 
If an evolution possesses a memory, the time derivative has to be replaced by an operator, for example, an integral expression: $B \psi = - i \hbar c \mbox{\boldmath$\alpha$} \cdot \mbox{\boldmath$\nabla$} \psi + \beta m c^2 \psi$ with, for example, $B \psi = \int B(t, t^\prime) \psi(t^\prime) dt^\prime$. 
In the case that the kernel $B(t, t^\prime)$ of that kind of equation possesses certain properties (it should be analytic), then one can expand that kernel resulting in a system of partial differential equations with an infinite sum of terms containing time derivatives of arbitrary order, $\sum_0^\infty a_i \partial_t^i \psi$. 

Therefore, if the quantum field does not evolve uniquely deterministic or if quantum theory has a memory, then the resulting field equation in these cases contains arbitrary high orders of time derivatives. 
In a first approximation, this may be modeled by adding to a conventional quantum field equation like the Dirac equation or the Schr\"odinger equation, a term containing a second time derivative: 
\begin{equation}
i \hbar \frac{\partial}{\partial t} \psi = - i \hbar c \mbox{\boldmath$\alpha$} \cdot \mbox{\boldmath$\nabla$} \psi + \beta m c^2 \psi - \epsilon \frac{\hbar^2}{m c^2} \frac{\partial^2}{\partial t^2} \psi \, .  \label{DiracMemory}
\end{equation}
In order to make $\epsilon$ dimensionless, we introduced a factor $1/m c^2$ in the term containing the last term. 
Here we assume that $\epsilon$ is constant, i.e.~does not depend on time or position. 
For $\epsilon = 0$, the above equation reduces to the usual Dirac equation. 
It is clear that the last term in (\ref{DiracMemory}) violates Lorentz covariance. 

This modified Dirac equation can be derived from a Lagrangian:
\begin{eqnarray}
{\cal L} & = & \frac{1}{2} i \hbar \left(\psi^+ \partial_t \psi - \partial_t \psi^+ \psi\right) \nonumber\\
& & + \frac{1}{2} i \hbar c \left(\psi^+ \mbox{\boldmath$\alpha$} \cdot \mbox{\boldmath$\nabla$} \psi - \mbox{\boldmath$\nabla$}\psi^+ \cdot \mbox{\boldmath$\alpha$} \psi\right) - m c^2 \psi^+ \beta \psi + \epsilon \frac{\hbar^2}{m c^2} \partial_t \psi^+ \partial_t\psi \\ 
& = & \frac{1}{2} i \hbar \left(\bar\psi \gamma^a \partial_a \psi - \partial_a\bar\psi \gamma^a \psi\right) - m c^2 \bar\psi \psi - \epsilon \frac{\hbar^2}{m c^2} \partial_t \bar\psi \gamma^0 \partial_t\psi \, . 
\end{eqnarray}
This implies, in particular, that we have a conservation law
\begin{equation}
0 = \partial_t \rho + \mbox{\boldmath$\nabla$} \cdot \mbox{\boldmath$j$}
\end{equation}
with the probability density
\begin{equation}
\rho = \psi^+ \psi - \epsilon \frac{\hbar^2}{m c^2} i \left(\partial_t \psi^+ \psi - \psi^+ \partial_t \psi\right)
\end{equation}
and a current
\begin{equation}
\mbox{\boldmath$j$} = \hbar c \psi^+ \mbox{\boldmath$\alpha$} \psi \, .
\end{equation}

The coupling to the electromagnetic field can, as usual, be introduced by means of the minimal coupling procedure:
\begin{eqnarray}
{\cal L} & = & \frac{1}{2} i \hbar \left(\psi^+ (\partial_t + i e \phi) \psi - (\partial_t - i e \phi) \psi^+ \psi\right) \nonumber\\
& & + \frac{1}{2} i \hbar c \left(\psi^+ \mbox{\boldmath$\alpha$} \cdot \left(\mbox{\boldmath$\nabla$} - \frac{i e}{\hbar c} \mbox{\boldmath$A$}\right) \psi - \left(\mbox{\boldmath$\nabla$} + \frac{i e}{\hbar c} \mbox{\boldmath$A$}\right)\psi^+ \cdot \mbox{\boldmath$\alpha$} \psi\right) \nonumber\\
& & - m c^2 \psi^+ \beta \psi + \epsilon \frac{\hbar^2}{m c^2} (\partial_t - i e \phi) \psi^+ (\partial_t + i e \phi)\psi \, .
\end{eqnarray}
The corresponding modified Dirac equation is
\begin{equation}
i \hbar \partial_t \psi = - i \hbar c \mbox{\boldmath$\alpha$} \cdot \left(\mbox{\boldmath$\nabla$} - \frac{i e}{\hbar c} \mbox{\boldmath$A$}\right) \psi + \beta m c^2 \psi + e \phi + \epsilon \frac{\hbar^2}{m c^2} (\partial_t - i e \phi)^2 \psi \, .
\end{equation}

\section{Plane wave solutions and neutrino propagation}\index{neutrino}

It is not difficult to present an exact plane wave solution for (\ref{DiracMemory}). 
Inserting the ansatz $\psi = \exp\left(i (E t - \mbox{\boldmath$p$} \cdot \mbox{\boldmath$x$})\right) a$ into (\ref{DiracMemory}) gives $Ea = \left(\mbox{\boldmath$\alpha$} \cdot \mbox{\boldmath$p$} c + \beta m c^2 + \frac{\epsilon}{m c^2} E^2\right) a$, or\footnote{Indices with a hat run from 1 to 3 and, for example, $p_{\hat a}$ is represented by $\mbox{\boldmath$p$}$.}
\begin{equation}
0 = \left(\gamma^0 \left(E - \frac{\epsilon}{m c^2} E^2\right) + \gamma^{\hat a} p_{\hat a} c - m c^2\right) a \, . \label{PlaneWaveCond}
\end{equation}
The corresponding dispersion relation reads
\begin{equation}
\left(E - \frac{\epsilon}{m c^2} E^2\right)^2 - {\mbox{\boldmath$p$}}^2 c^2 = m^2 c^4 
\end{equation}
which possesses the four solutions
\begin{equation}
\begin{split}
E^{(1)}_\pm & = \frac{m c^2 - \sqrt{m^2 c^4 \mp 4 \epsilon m c^3 \sqrt{{\mbox{\boldmath$p$}}^2 c^2 + m^2 c^2}}}{2 \epsilon}  \\ 
E^{(2)}_\pm & = \frac{m c^2 + \sqrt{m^2 c^4 \pm 4 \epsilon m c^3 \sqrt{{\mbox{\boldmath$p$}}^2 c^2 + m^2 c^2}}}{2 \epsilon} \, .
\end{split}
\end{equation}
For small $\epsilon$ this reduces to
\begin{equation}
\begin{split}
E^{(1)}_\pm & = \pm c \sqrt{{\mbox{\boldmath$p$}}^2 + m^2 c^2} + \epsilon \left(m c^2 + \frac{{\mbox{\boldmath$p$}}^2}{m}\right) \\ 
E^{(2)}_\pm & = \frac{m c^2}{\epsilon} \pm c \sqrt{{\mbox{\boldmath$p$}}^2 + m^2 c^2} - \epsilon \left(m c^2 + \frac{{\mbox{\boldmath$p$}}^2}{m}\right) \, .
\end{split}
\end{equation}
That means that for $\epsilon \rightarrow 0$ the solutions $E^{(1)}_\pm$ reduce to the well--known solutions. 
The other two solutions are new and diverge for small $\epsilon$. However, this large quantity will drop out by considering energy differences. 
The $\pm$ in the solutions for the energy corresponds to positive/negative energies, the (1), (2) corresponds to the two solutions which come up with $\epsilon \neq 0$.

In the high energy limit $m \rightarrow 0$ we get ($p = |{\mbox{\boldmath$p$}}|$)
\begin{equation}
\begin{split}
E^{(1)}_\pm & = \pm c p \pm \frac{m^2 c^3}{2 p} + \epsilon \left(m c^2 + \frac{p^2}{m}\right) \\
E^{(2)}_\pm & = \frac{m c^2}{\epsilon} \pm c p \pm \frac{m^2 c^3}{2 p} - \epsilon \left(m c^2 + \frac{{\mbox{\boldmath$p$}}^2}{m}\right) \, .
\end{split}
\end{equation}
In the low energy limit $\mbox{\boldmath$p$} \rightarrow 0$ we get instead
\begin{equation}\label{energyp}
\begin{split}
E^{(1)}_\pm & = \pm m c^2 \pm \frac{{\mbox{\boldmath$p$}}^2}{2 m} + \epsilon \left(m c^2 + \frac{{\mbox{\boldmath$p$}}^2}{m}\right) \\
E^{(2)}_\pm & = \frac{m c^2}{\epsilon} \pm m c^2 \pm \frac{{\mbox{\boldmath$p$}}^2}{2 m} - \epsilon \left(m c^2 + \frac{{\mbox{\boldmath$p$}}^2}{m}\right) \, .
\end{split}
\end{equation}

The exact expression for the group velocity reads
\begin{equation}\label{groupexact}
\begin{split}
(v^{(1)}_\pm)^{\hat a} & = \pm \frac{p_{\hat a} c}{\sqrt{{\mbox{\boldmath$p$}}^2 + m^2 c^2} \sqrt{1 \mp 4 \frac{\epsilon}{m c} \sqrt{{\mbox{\boldmath$p$}}^2 + m^2 c^2}}} \\
(v^{(2)}_\pm)^{\hat a} & = \pm \frac{p_{\hat a} c}{\sqrt{{\mbox{\boldmath$p$}}^2 + m^2 c^2} \sqrt{1 \pm 4 \frac{\epsilon}{m c} \sqrt{{\mbox{\boldmath$p$}}^2 + m^2 c^2}}} \, .
\end{split}
\end{equation}
Therefore, for positive energies, particles with the same momentum but with different directions propagate with different velocities (the same is true for particles with negative energies). 
This property can be used for a comparison with data from neutrino propagation. 

For small $\epsilon$ we get from (\ref{groupexact}) 
\begin{equation}
\begin{split}
(v^{(1)}_\pm)^{\hat a} & = \pm \frac{p_{\hat a} c}{\sqrt{{\mbox{\boldmath$p$}}^2 + m^2 c^2}} + 2 \epsilon \frac{p_{\hat a}}{m} \\
(v^{(2)}_\pm)^{\hat a} & = \pm \frac{p_{\hat a} c}{\sqrt{{\mbox{\boldmath$p$}}^2 + m^2 c^2}} - 2 \epsilon \frac{p_{\hat a}}{m} \, ,
\end{split}
\end{equation}
for $m \rightarrow 0$
\begin{equation}\label{groupformapprox0}
\begin{split}
(v^{(1)}_\pm)^{\hat a} & = \pm \frac{p_{\hat a}}{p} c \mp \frac{m^2 c^3}{2 p^3} p_{\hat a} + 2 \epsilon \frac{p_{\hat a}}{m} \\
(v^{(2)}_\pm)^{\hat a} & = \pm \frac{p_{\hat a}}{p} c \mp \frac{m^2 c^3}{2 p^3} p_{\hat a} - 2 \epsilon \frac{p_{\hat a}}{m} \, ,
\end{split}
\end{equation}
and for $p_{\hat a} \rightarrow 0$
\begin{equation}
\begin{split}
(v^{(1)}_\pm)^{\hat a} & = \pm \frac{p_{\hat a}}{m} + 2 \epsilon \frac{p_{\hat a}}{m} \\
(v^{(2)}_\pm)^{\hat a} & = \pm \frac{p_{\hat a}}{m} - 2 \epsilon \frac{p_{\hat a}}{m} \, .
\end{split}
\end{equation}

For a comparison with data from the propagation of neutrinos, which may be produced in connection with GRBs, we use (\ref{groupformapprox0}). 
We compare the arrival time of neutrinos with the arrival time of light over a distance of $l = 10^{10}\;\hbox{ly}$. 
If the neutrinos and the photons are produced during the same event, and if we take the mass of the neutrinos to be 1 eV and the momentum $p = 10^5\;\hbox{GeV}$, then we get as difference of the time--of--arrival
\begin{equation}
\Delta t = \frac{l}{c} - \frac{l}{v_+^{(1,2)}} \approx \frac{l}{c} \left(\frac{m^4 c^4}{8 p^4} \pm 2 \epsilon \frac{p}{m c}\right) \approx \left(4 \times 10^{-9} \pm 6.4 \, \epsilon \, 10^{31}\right)\;\hbox{sec} \, .
\end{equation}
The first term can be neglected compared to the second one, so that we get
$|\Delta t| = |\epsilon| 6.4 \times 10^{31}\;\hbox{sec}$. 
Asssuming a temporal structure of the source of about a millisecond \cite{Bhatetal92} and {\it assuming a null--result}, then we can get from observations of the propagation of neutrinos and of photons the estimate
\begin{equation}
|\epsilon| \leq 1.6 \times 10^{- 35} \, . \label{EstEpsNeutrino}
\end{equation}
Thus, neutrino observations in the future have the potentiality of high precision determination of the parameter $\epsilon$. 
Any $|\epsilon|$ larger than that given by (\ref{EstEpsNeutrino}) should be detectable by this means. 

In quantum gravity theories, $|\epsilon|$ is proportional to the ratio of the Planck length and some intermediate length, $\epsilon = \kappa l_p/L$ where $\kappa$ is assumed to be of the order 1 \cite{BL_AlfraoMoralesTecotlUrrutia00}.  
If we take $L = \hbar/p$, then, in terms of $\kappa$, the above estimate means $|\kappa| \leq 1.6 \times 10^{-21}$ which certainly is in contradiction to the assumption that  $\kappa$ is of the order 1. 
From this we conclude, that, if the Dirac equation contains an additional quantum gravity induced term with the second time derivative of the neutrino field, then this term should be observable in the future by comparing neutrino propagation with photon propagation. 

However, from the derivation of the modifications of the Dirac equation \cite{BL_AlfraoMoralesTecotlUrrutia00}, the parameter $\epsilon$ or $\kappa$ may be constant over the scale $L$ of the ``weave'' states only. 
Therefore, it may be possible that the effect, as we calculated it, may not occur due to an averaging to zero over larger distances. 
Consequently, it is necessary also to perform tests which take place on a small scale only. 
One kind of such tests is atomic spectroscopy what we are going to discuss below.  

\section{The non--relativistic limit}

\subsection{The non--relativistic field equation}

First we calculate the modified Pauli equation corresponding to eqn (\ref{DiracMemory}). 
In order to do so, we first subtract the rest energy from the wave function by means of the substitution
\begin{equation}
\psi = e^{- \frac{i}{\hbar} (1 + \epsilon) m c^2 t} \psi^\prime  \label{Eq:FWAnsatz}
\end{equation}
resulting in an elimination of the rest mass in one part of the wave function. 
This gives
\begin{eqnarray}
(1 + 2 \epsilon (1 + \epsilon)) i \hbar \partial_t \psi^\prime & = & - i \hbar c \mbox{\boldmath$\alpha$} \cdot \left(\mbox{\boldmath$\nabla$} - \frac{i e}{\hbar c} \mbox{\boldmath$A$}\right) \psi^\prime + (\beta - 1) m c^2 \psi^\prime + e \phi \psi^\prime \nonumber\\
& & - \epsilon \frac{\hbar^2}{m c^2} (\partial_t + i e \phi)^2 \psi^\prime \, .
\end{eqnarray}
With the projection operators $P_\pm := \frac{1}{2} (1 \pm \beta)$ we define the `large' and `small' parts of the wave function: $\psi^\prime_\pm = P_\pm \psi^\prime$. 
Multiplication of (\ref{DiracMemory}) with $P_+$ and $P_-$ gives the two equations
\begin{align}
(1 + 2 \epsilon (1 +€\epsilon)) i \hbar \partial_t \psi^\prime_+ & = - i \hbar c \mbox{\boldmath$\alpha$} \cdot \!\left(\mbox{\boldmath$\nabla$} - \frac{i e}{\hbar c} \mbox{\boldmath$A$}\right)\!\psi^\prime_- + e \phi \psi_+ - \epsilon \frac{\hbar^2}{m c^2} (\partial_t + i e \phi)^2 \psi^\prime_+ \label{nonrel1} \\
(1 + 2 \epsilon) i \hbar \partial_t \psi^\prime_- & = - i \hbar c \mbox{\boldmath$\alpha$} \cdot \left(\mbox{\boldmath$\nabla$} - \frac{i e}{\hbar c} \mbox{\boldmath$A$}\right) \psi^\prime_+ + e \phi \psi_- - 2 m c^2 \psi^\prime_-  \nonumber\\
& \quad - \epsilon \frac{\hbar^2}{m c^2} (\partial_t + i e \phi)^2 \psi^\prime_- \, .
\end{align}
As usual, we assume that in the second equation the energies $i \hbar \partial_t \psi^\prime_-$ and $e \phi \psi_-$ are small compared with the rest mass term $m c^2 \psi^\prime_-$. 
Therefore, we approximate
\begin{equation}
\psi^\prime_- \approx - \frac{1}{2 m c} i \hbar \mbox{\boldmath$\alpha$} \cdot \left(\mbox{\boldmath$\nabla$} - \frac{i e}{\hbar c} \mbox{\boldmath$A$}\right) \psi^\prime_+  \, . 
\end{equation}
Inserting this into the first equation (\ref{nonrel1}) gives
\begin{eqnarray}
(1 + 2 \epsilon (1 + \epsilon)) i \hbar \partial_t \psi^\prime_+ & = & - \frac{\hbar^2}{2 m} \left(\mbox{\boldmath$\nabla$} - \frac{i e}{\hbar c} \mbox{\boldmath$A$}\right)^2 \psi^\prime_+ + e \phi \psi^\prime_+ \nonumber\\
& & + \frac{e}{m c} \mbox{\boldmath$\Sigma$} \cdot \mbox{\boldmath$B$} - \epsilon \frac{\hbar^2}{m c^2} (\partial_t + i e \phi)^2 \psi^\prime_+ \, ,
\end{eqnarray}
with $\mbox{\boldmath$\Sigma$} = \begin{pmatrix}\mbox{\boldmath$\sigma$} & 0 \\ 0 & \mbox{\boldmath$\sigma$}\end{pmatrix}$ and $\mbox{\boldmath$B$} = \mbox{\boldmath$\nabla$} \times \mbox{\boldmath$A$}$. 
Here $\mbox{\boldmath$\sigma$}$ are the three Pauli matrices. 
After division through $1 + 2 \epsilon (1 + \epsilon)$ we get 
\begin{equation}
i \hbar \partial_t \psi^\prime_+ = - \frac{\hbar^2}{2 m^*} \left(\mbox{\boldmath$\nabla$} - \frac{i e}{\hbar c} \mbox{\boldmath$A$}\right)^2 \psi^\prime_+ + e^* \phi \psi^\prime_+ + \frac{e}{m^* c} \mbox{\boldmath$\Sigma$} \cdot \mbox{\boldmath$B$} - \epsilon \frac{\hbar^2}{m^* c^2} (\partial_t + i e \phi)^2 \psi^\prime_+ \, ,
\end{equation}
where we absorbed the $\epsilon$--factor in a redefinition of mass and charge: $m^* := (1 + 2 \epsilon (1 + \epsilon)) m$, $e^* = e/(1 + 2 \epsilon (1 + \epsilon))$.

As our result we get a modified Schr\"odinger equation for a two--spinor $\psi$
\begin{equation}
i \hbar \frac{\partial}{\partial t} \psi = - \frac{\hbar^2}{2 m^*}
\left(\mbox{\boldmath$\nabla$} - \frac{i e}{\hbar c} \mbox{\boldmath$A$}\right)^2 \psi + e^* \phi \psi + \frac{e}{m c} \mbox{\boldmath$\sigma$} \cdot \mbox{\boldmath$B$} - \epsilon \frac{\hbar^2}{m^* c^2} (\partial_t + i e \phi)^2 \psi \, . \label{SchroedingerMemory}
\end{equation}
This equation can also be derived from a Lagrangian
\begin{eqnarray}
{\cal L} & = & \frac{i}{2} \hbar (\psi^* (\partial_t - i e \phi) \psi - (\partial_t + i e \phi) \psi^* \psi) \\ 
& & - \frac{\hbar^2}{2 m^*} \left(\mbox{\boldmath$\nabla$} + \frac{i e}{\hbar c} \mbox{\boldmath$A$}\right) \psi^* \cdot \left(\mbox{\boldmath$\nabla$} - \frac{i e}{\hbar c} \mbox{\boldmath$A$}\right) \psi - \epsilon \frac{\hbar^2}{m^* c^2} (\partial_t - i e \phi) \psi^* (\partial_t + i e \phi) \psi \, . \nonumber
\end{eqnarray}

Equation (\ref{SchroedingerMemory}) also possesses plane wave solutions whose energies are given by (\ref{energyp}).
We also have a conservation law 
\begin{equation}
\frac{d}{dt}\rho + \mbox{\boldmath$\nabla$} \mbox{\boldmath$j$} = 0
\end{equation}
with the probability
\begin{equation}
\rho = \psi^* \psi + i \epsilon \frac{\hbar}{m^* c^2} \left(\dfrac{\partial}{\partial t} \psi^* \psi - \psi^* \dfrac{\partial}{\partial t} \psi\right) \, .
\end{equation}
and the current
\begin{equation}
\mbox{\boldmath$j$} = \frac{i \hbar}{2 m^*}
\left(\nabla \psi^* \psi - \psi^* \nabla \psi\right) \, .
\end{equation}
Like in the Klein--Gordon equation there seems to exist the possibility to get negative probablities. 
However, using the Schr\"odinger equation, we get for the probability
\begin{eqnarray}
\rho & = & \psi^* \psi - \epsilon \frac{1}{m c^2} \left(- i \hbar \dfrac{\partial}{\partial t} \psi^* \psi + \psi^* i \hbar \dfrac{\partial}{\partial t} \psi\right) \nonumber\\
& = & \psi^* \psi + \epsilon \frac{1}{m^* c^2} \left(\frac{\hbar^2}{2 m^*}
\Delta \psi \psi + \psi^* \frac{\hbar^2}{2 m^*} \Delta \psi\right) \, .
\end{eqnarray}
This quantity is strictly positive if $\epsilon < 0$, and for $\epsilon > 0$ this is positive if $|\epsilon|$ is small enough. 

\subsection{Modifications of the energy levels}

It is possible to calculate the energy levels of the hydrogen atom \index{hydrogen atom} exactly. 
In order to do so we choose $\mbox{\boldmath$A$} = 0$ and $\phi = \phi(r)$ as the usual spherically symmetric electrostatic potential:
\begin{equation}
i \hbar \frac{\partial}{\partial t} \psi = - \frac{\hbar^2}{2 m^*}
\Delta \psi + e^* \phi \psi + \epsilon \frac{1}{m^* c^2}
\left(i \hbar \frac{\partial}{\partial t} - e \phi\right)^2 \psi \, .
\end{equation}
We asume a stationary solution, then $i \hbar \partial_t \psi = E \psi$, and we get from the above Hamiltonian
\begin{equation}
\left(E - \epsilon \frac{1}{m^* c^2} E^2\right) \psi = - \frac{\hbar^2}{2 m^*} \Delta \psi 
+ \left(e^* - 2 \epsilon \frac{e}{m^* c^2} E\right) \phi \psi + \epsilon \frac{1}{m^* c^2} e^2 \phi^2 \psi \, .
\end{equation}
With $\Delta = \frac{1}{r^2} \frac{\partial}{\partial r} \left(r^2 \frac{\partial}{\partial r}\right) - \frac{1}{r^2} {\widehat{\mbox{\boldmath$L$}}}^2$, where $\widehat{\mbox{\boldmath$L$}}$ is the angular momentum operator with the eigenvalue equation ${\widehat{\mbox{\boldmath$L$}}}^2 Y_l^m = l (l + 1) Y_l^m$ with $l = 1, 2, 3, \ldots$. 
With a splitting of the wave function into a radial and an angular part $\psi = R(r) Y_l^m(\vartheta, \varphi)$ we get the radial part of the wave equation
\begin{eqnarray}
\left(E - \epsilon \frac{1}{m^* c^2} E^2\right) R & = & - \frac{\hbar^2}{2m^*} \left(\frac{1}{r^2} \frac{\partial}{\partial r} \left(r^2 \frac{\partial}{\partial r} R\right) - \frac{1}{r^2} l (l + 1) R\right) \nonumber\\
& & + \left(e^* - 2 \epsilon \frac{e}{m^* c^2} E\right) \phi R  + \epsilon \frac{1}{m^* c^2} e^2 \phi^2 R \, .
\end{eqnarray}
With the explicit expression for the electrostatic potential $\phi = - e/r$ we get by multiplication with $2 m^*/\hbar^2$ 
\begin{eqnarray}
0 & = & \frac{d^2}{d r^2} R + \frac{2}{r} \frac{d}{d r} R + \frac{2 m^*}{\hbar^2}\left(E - \epsilon \frac{1}{m^* c^2} E^2\right) R \nonumber\\
& & + \frac{2 m^*}{\hbar^2} \left(e^* - 2 \epsilon \frac{e}{m^* c^2} E\right) \frac{e}{r} R - \frac{1}{r^2} \left(l (l + 1) + \epsilon \frac{2 e^4}{\hbar^2 c^2} \right) R \, .
\end{eqnarray}
Since the $r$--dependence is the same as for the usual hydrogen atom, this equation can be solved using the standard scheme: 
With 
\begin{equation} \label{Eq:abbrev}
{\cal E} := \frac{2 m^*}{\hbar^2} \left(E - \frac{\epsilon}{m^* c^2} E^2\right), \quad q := \frac{2 m^* e}{\hbar^2} \left(e^* - \frac{2 \epsilon e}{m^* c^2} E\right), \quad \ell := l (l + 1) + 2 \epsilon \alpha^2
\end{equation}
where $\alpha = e^2/\hbar c$ is the fine structure constant, 
we get for the radial part of the wave function 
\begin{equation}
\left(\frac{d^2}{d r^2} + \frac{2}{r} \frac{d}{d r} + {\cal E} + \frac{q}{r} - \frac{\ell}{r^2} \right) R = 0\, .
\end{equation}
We introduce dimensionless coordinates by
\begin{equation}
r^{\prime}=\frac{r}{r_0}, \quad - {\cal E} = \frac{1}{4 r_0^2}, 
\quad q^\prime = q r_0 \label{Definitions2}
\end{equation}
and get with $R^\prime(r^\prime) := R(r(r^\prime))$
\begin{equation}
\left(\frac{d^2}{d {r^\prime}^2} + \frac{2}{r^\prime} \frac{d}{d {r^\prime}} - \frac{1}{4} + \frac{q^\prime}{r^\prime} - \frac{1}{{r^\prime}^2} \ell\right) R^\prime = 0\, .
\end{equation}
We also introduce a new variable $f(r^\prime)$ through
\begin{equation}\label{BL_equ:ansatz}
R = e^{- \frac{1}{2} r^\prime} r^{\prime\gamma} f(r^\prime)
\end{equation}
with a parameter $\gamma$ which will be specified later. 
We get the following equation for the function $f$:
\begin{equation}\label{equ:allg}
0 = r \frac{d^2 f(r^\prime)}{d{r^\prime}^2}  + \left(2 \gamma + 2 - r\right) \frac{df(r^\prime)}{dr^\prime} + \left(\frac{\gamma (\gamma + 1)- \ell}{r^{\prime}}+ q^{\prime} - \gamma - 1\right) f(r^\prime)\;.
\end{equation}
In order to solve this equation we specify the value of $\gamma$ by the requirement that the term $\sim 1/r^\prime$ should vanish:
\begin{equation}
\gamma \left(\gamma + 1\right) - \ell = 0 \; .
\end{equation}
This gives the two possibilities
\begin{equation}
\gamma_\pm = - \frac{1}{2} \pm \sqrt{\ell + \frac{1}{4}} = - \frac{1}{2} \pm \sqrt{\left(l + \frac{1}{2}\right)^2 + 2 \epsilon \alpha^2} \, ,
\end{equation}
and from (\ref{equ:allg}) the differential equations
\begin{equation} \label{confl}
z f^{\prime\prime} + (\vartheta_\pm - z) f^{\prime} - \beta_\pm f = 0 \, ,
\end{equation}
with
\begin{eqnarray}
\vartheta_\pm & = & 2 \gamma_\pm + 2 = 1 \pm 2 \sqrt{\ell + \frac{1}{4}}  \\
\beta_\pm & = & \gamma_\pm + 1 - q^{\prime} = \frac{1}{2} \pm \sqrt{\ell + \frac{1}{4}} - q^{\prime} \, .\label{Eq:beta}
\end{eqnarray}
Eqn.\ (\ref{confl}) is the confluent hypergeometric differential equation
with the solution \cite{GradshteynRyzhik83}
\begin{equation}
f (\beta_\pm,\vartheta_\pm, z) = \sum_{\nu=0}^{\infty} \frac{(\beta_\pm + \nu)! \vartheta_\pm!}{\beta_\pm! (\vartheta_\pm + \nu)!}\frac{z^{\nu}}{\nu!} \; . \label{CHS}
\end{equation}
which is appropriate for our problem.

It is clear that, in order to get no infinite terms, $\vartheta_\pm$ is not allowed to be a negative integer: $\vartheta_\pm \neq -1, -2, \ldots$, which is fulfilled if $\epsilon \neq 0$ and $|\epsilon| < 1$.
For $\epsilon = 0$ we cannot use the solution $\vartheta_-$.
In addition, if the sum in (\ref{CHS}) does not terminate, then the solution diverges for large $r$ faster than $\exp\left(\frac{1}{2} r^\prime\right)$ which leads to non--normalizable solutions.
The condition for a termination of the sum is $\beta_\pm \in \mathbb{Z}^-$, or
\begin{equation}
\beta_\pm = -k, \quad\quad\quad k \in \mathbb{N}\;.
\end{equation}
With (\ref{Eq:beta}), (\ref{Definitions2}), and (\ref{Eq:abbrev}) we get four energy eigenvalues $E$
\begin{eqnarray}
E_\pm^{(1)} & = & \frac{m^* c^2}{2 \epsilon} \frac{K_\pm^2 - 2 \epsilon \alpha \alpha^* \frac{e}{e^*}}{K_\pm^2 - 2 \epsilon \alpha^2} + \frac{m^* c^2}{2 \epsilon} \frac{K_\pm \sqrt{K_\pm^2 + 2 \epsilon \alpha \alpha^* (1 - 2 \frac{e}{e^*})}}{K_\pm^2 - 2 \epsilon \alpha^2} \\
E_\pm^{(2)} & = & \frac{m^* c^2}{2 \epsilon} \frac{K_\pm^2 - 2 \epsilon \alpha \alpha^* \frac{e}{e^*}}{K_\pm^2 - 2 \epsilon \alpha^2} - \frac{m^* c^2}{2 \epsilon} \frac{K_\pm \sqrt{K_\pm^2 + 2 \epsilon \alpha \alpha^* (1 - 2 \frac{e}{e^*})}}{K_\pm^2 - 2 \epsilon \alpha^2} 
\end{eqnarray}
with $K_\pm = k - \frac{1}{2} \mp \sqrt{(l + \frac{1}{2})^2 + 2 \epsilon \alpha^2}$ and $\alpha^* = {e^*}^2/\hbar c$. 
To first order in $\epsilon$ we get
\begin{eqnarray}
E_+^{(1)} & = & m^* c^2 \left(\frac{1}{\epsilon} + \frac{\alpha^2}{2} \frac{1}{(1 - k + l)^2} - \alpha^4 \epsilon \frac{5 - 8 k + 2 l}{4 (1 - k + l)^4 (1 + 2 l)} + {\cal O}(\epsilon^2)\right) \label{Energy1+} \\
E_-^{(1)} & = & m^* c^2 \left(\frac{1}{\epsilon} + \frac{\alpha^2}{2} \frac{1}{(k + l)^2} + \epsilon \alpha^4 \frac{3 - 8 k - 2 l}{4 (k + l)^4 (1 + 2 l)} + {\cal O}(\epsilon^2)\right) \label{Energy1-} \\
E_+^{(2)} & = & - m^* c^2 \left(\frac{\alpha^2}{2} \frac{1}{(1 - k + l)^2} - \epsilon \alpha^4 \frac{5 - 8 k + 2 l}{4 (1 - k + l)^4 (1 + 2 l)} + {\cal O}(\epsilon^2)\right) \label{Energy2+}\\
E_-^{(2)} & = & - m^* c^2 \left(\frac{\alpha^2}{2} \frac{1}{(k + l)^2} + \epsilon \alpha^4 \frac{3 - 8 k - 2 l}{4 (k + l)^4 (1 + 2 l)} + {\cal O}(\epsilon^2)\right) \, . \label{Energy2-}
\end{eqnarray}
The solution is then given by
\begin{equation}
R(r) = R(r(r^\prime)) = R^\prime(r^\prime) = A e^{- \frac{1}{2} r^\prime} r^{\prime\gamma} f(r^\prime) = A e^{- \frac{1}{2} r^\prime} {r^\prime}^{\gamma_\pm} f(-k, \vartheta_\pm, r^\prime)
\end{equation}
where $A$ is a normalization constant and where all parameters depend on $l$.
 
For $\epsilon \neq 0$ all these energy values are well defined. 
Even for very small $\epsilon$ the first two energies (\ref{Energy1+},\ref{Energy1-}) are valuable solutions, too, since only energy differences are observable and thus the first term drops out. 
However, there are two reasons which justify to drop the first two solutions: (i) Except the first term $mc^2/\epsilon$, the upper two sets of energy levels (\ref{Energy1+},\ref{Energy1-}) are the same as the lower two sets (\ref{Energy2+},\ref{Energy2-}) up to sign. For the first two sets of energy levels the continuum is below the discrete spectrum. By postulating that all particles fall into the lowest energy level, then all atoms will fall into the continuous part of the spectrum which has never been observed. 
(One also can postulate that all particles want to go to the highest energy level. But due to the symmetry of the two sets of spectra, this will give the same answer. Therefore we do not consider the upper two sets by convention.) 
(ii) We want to describe small modifications of the known energy levels given for $\epsilon = 0$. 
Owing to these reasons, we keep the last two sets of energy levels. 

Defining for $E_+^{(2)}$ the principal quantum number $n := 1 - k + l$, then the energy levels are given by 
\begin{equation}
E_+^{(2)} = - m^* c^2 \frac{\alpha^2}{2} \left(\frac{1}{n^2} + \epsilon \alpha^2 \left(\frac{3}{2 n^4} - \frac{4}{n^3 (1 + 2 l)}\right) + {\cal O}(\epsilon^2)\right) \label{Energy1}
\end{equation}
and if we define for $E_-^{(2)}$ the principal quantum number $n := k + l$, then the energy levels are
\begin{equation}
E_-^{(2)} = - m^* c^2 \frac{\alpha^2}{2} \left(\frac{1}{n^2} + \epsilon \alpha^2 \left(\frac{3}{2 n^4} - \frac{4}{n^3 (1 + 2 l)}\right) + {\cal O}(\epsilon^2)\right) \, ,
\end{equation}
where $m c^2 \alpha^2/2$ is the Rydberg constant $Ry$. 
It is remarkable, that in both cases we get the same scheme of energy differences. 
Therefore, at this order we cannot distinguish between energies $E_+$ and $E_-$. 

There are two differences between this result and the usual spectrum of the  hydrogen atom: (i) No degeneracy with respect to $l$. (ii) Additional $1/n^4$-- and $1/n^3$--terms. 
From (\ref{Energy1}) we get for the Lyman--series with $n = 1$ and $l = 0$ as ground state, for example, the energy differences
\begin{equation}
\Delta E_{\hbox{\scriptsize Lyman}} = Ry \left(\! 1 - \frac{1}{n^2} + \epsilon \alpha^2 \left(\frac{3}{2} \left(1 - \frac{1}{n^4}\right) - 4 \left(1 - \frac{1}{n^3 (1 + 2 l)}\right)\right)  + {\cal O}(\epsilon^2)\right).
\end{equation}
The additional terms $1/n^4$ and $n^3 (1 + 2 l)$ modify the structure of this series. 
The ionization energy is given by $n \rightarrow \infty$ and is
\begin{equation}
E_{\hbox{\scriptsize ionization}} = Ry \left(1 - \frac{5}{2} \epsilon \alpha^2 + {\cal O}(\epsilon^2)\right) \, .
\end{equation}
From numbers related to $\Delta E_{\hbox{\scriptsize Lyman}}$ or $E_{\hbox{\scriptsize ionization}}$ one can draw estimates on the value of $\epsilon$. 

Since the accuracy $\delta \Delta E/\Delta E$ for recent measurements, see for example \cite{Udemetal97}, is of the order of $10^{-13}$ which agrees completely with the conventional theory, we can conclude that the relative deviation from the usually calculated energy has to be smaller than this uncertainty:
\begin{equation}
\frac{\delta\Delta E}{\Delta E} = \epsilon \alpha^2 \frac{\left|- \frac{5}{2} - \frac{3}{2 n^4} + \frac{4}{n^3}\right|}{1 - \frac{1}{n^2}} \leq 10^{-13} \, .
\end{equation} 
This means that
\begin{equation}
\epsilon \leq 10^{-13} \frac{1}{\alpha^2} \frac{1 - \frac{1}{n^2}}{\left|- \frac{5}{2} - \frac{3}{2 n^4} + \frac{4}{n^3}\right|} = 10^{-13} \frac{1}{\alpha^2}  \frac{3}{8 + \frac{3}{8}} \approx 7 \times 10^{-10}\, .
\end{equation}
In terms of a $\kappa$ as introduced after Eqn.~(\ref{EstEpsNeutrino}) we have the estimate $\kappa \leq 7\times 10^4$ which is outside the assumption that $\kappa$ is of the order unity. Therefore, the accuracy of atomic spectroscopy is still at least five orders of magnitude too small in order to be able to detect any influence of quantum gravity on atomic levels.

\section{Conclusion}

We have shown that an additional term in the Dirac equation containing a second time derivative, as it is motivated from quantum gravity, influences neutrino propagation and atomic spectroscopy. 
While neutrino propagation may be capable to `see' this additional term, the accuracy of spectroscopy has still to be improved by five orders of magnitude in order to be sensitive to this term. 

A clear difficulty with the hydrogen spectrum is that, precisely, it cannot be calculated exactly with an arbitrary accuracy with the present state--of--the--art. 
There is a number of corrections to the Dirac solutions: recoil, QED, finite nucleus size, see \cite{Pachuckietal96} for a review. 
These corrections scale with $1/n^3$ and amount to a few kilohertz. 
They are very difficult to calculate and there are still some discrepancies between theoretical results. Fortunately, some combinations of frequencies are independent of these corrections at their leading order and can be used for higher accuracies. 
Concerning the measurements themselves, there is presently a very rapid evolution towards much higher accuracies. For example, it was shown recently that, because of the extreme regularity of the frequency comb generated by femtosecond  lasers over a very wide spectrum \cite{Udemetal99}, they could be used to compare frequencies of oscillators which differ by several orders of magnitude. 
This provides a way to compare many transition frequencies of the hydrogen atom between themselves and with microwave clocks, with the potential accuracy of the cesium fountain clock which is presently $10^{-15}$ and should improve quickly by another order of magnitude. Also, the techniques to interrogate narrow transitions of cold atom hydrogen by sub--Doppler methods or atom interferometry have improved very significantly either in cold thermal beams (H\"ansch and coworkers in Garching, Biraben and coworkers in Paris) or in clouds generated from Bose--Einstein condensates (Kleppner and coworkers at MIT). A subkilohertz linewidth is presently achieved for the 1S-2S two--photon transition \cite{Udemetal99} and could still be reduced by one or two orders of magnitude in the near future. The hydrogen atom is thus potentially a universal clock by itself covering the full spectrum from UV to microwaves (hydrogen maser).

\section*{Acknowledgement}

Ch.B. thanks C.\ L\"ammerzahl and J.\ Mlynek for their hospitality at the University of Konstanz. 
We acknowledge financial support of the Optik--Zentrum of the University of Konstanz.


\begin{thebibliography}{8.}
\addcontentsline{toc}{section}{References}

\bibitem{FischbachTalmadge99}
E.~Fischbach and C.L. Talmadge.
\newblock {\em The Search for Non--{N}ewtonian Gravity}.
\newblock Springer--Verlag, New York, 1999.

\bibitem{BL_AlfraoMoralesTecotlUrrutia00}
J.~Alfaro, H.A. Morales-Tecotl, and L.F. Urrutia.
\newblock Quantum gravity corrections to neutrino propagation.
\newblock {\em Phys.\ Rev.\ Lett.}, 84:to appear, 2000.

\bibitem{BL_Ellisetal99}
J.~Ellis, N.E. Mavromatos, D.V. Nanopoulos, and G.~Volkov.
\newblock Gravitational--recoil effects on fermion propagation in space-time
  foam.
\newblock gr-qc/9911055.

\bibitem{AL93}
J.~Audretsch and C.~L{\"a}mmerzahl.
\newblock A new constructive axiomatic scheme for the geometry of space--time.
\newblock In Majer U. and Schmidt H.--J., editors, {\em Semantical Aspects of
  Space--Time Geometry}, page~21. BI Verlag, Mannheim, 1993.

\bibitem{LaemmerzahlBleyer99}
C.~L{\"a}mmerzahl and U.~Bleyer.
\newblock A quantum test theory for basic principles of special and general
  relativity with applications to atomic interferometry and Hughes--Drever type
  experiments, 1999.
\newblock preprint, University of Konstanz.

\bibitem{Shimony79}
A.~Shimony.
\newblock Proposed neutron interferometer test of some nonlinear variants of
  wave mechanics.
\newblock {\em Phys.\ Rev.}, A 20:394, 1979.

\bibitem{Weinberg89}
S.~Weinberg.
\newblock Testing quantum mechanics.
\newblock {\em Ann.\ Phys.}, 194:336, 1989.

\bibitem{Shulletal80}
C.G. Shull, D.K. Atwood, J.~Arthur, and M.A. Horne.
\newblock Search for a nonlinear variant of the {S}chr{\"o}dinger equation by
  neutron interferometry.
\newblock {\em Phys.\ Rev.\ Lett.}, 44:765, 1980.

\bibitem{Ellisetal84}
J.~Ellis, S.~Hagelin, D.V. Nanopoulos, and M.~Srednicki.
\newblock Search for violations of quantum mechanics.
\newblock {\em Nucl.\ Phys.}, B 241:381, 1984.

\bibitem{LB_Laemmerzahl98}
C.~L\"ammerzahl.
\newblock Quantum Tests of Foundations of General Relativity.
\newblock {\em Class.\ Quantum Grav.}, 14:13, 1998.

\bibitem{BurgbacherLaemmerzahlMacias99}
F.~Burgbacher, C.~L{\"a}mmerzahl, and A.~Macias.
\newblock Is there a stable hydrogen atom in higher dimensions?
\newblock {\em J.\ Math.\ Phys.}, 40:625, 1999.

\bibitem{Bhatetal92}
P.N. Bhat, G.J. Fishman, C.A. Meegan, R.B. Wilson, M.N. Brock, and W.S.
  Paciesas.
\newblock Evidence for sub--millisecond structure in a {$\gamma$}--ray burst.
\newblock {\em Nature}, 359:217, 1992.

\bibitem{GradshteynRyzhik83}
I.S. Gradshteyn and I.M. Ryzhik.
\newblock {\em Table of Integrals, Series, and Products}.
\newblock Academic Press, Orlando, 1983.

\bibitem{Udemetal97}
Th. Udem, A.~Huber, B.~Gross, J.~Reichert, M.~Prevedelli, M.~Weitz, and T.W.
  H{\"a}nsch.
\newblock Phase--coherent measurement of the {H}ydrogen 1s--2s transition
  frequency with an optical frequency interval divider chain.
\newblock {\em Phys.\ Rev.\ Lett.}, 79:2646, 1997.

\bibitem{KieferSingh91}
C.~Kiefer and T.P. Singh.
\newblock Quantum gravitational corrections to the functional {S}chr{\"o}dinger
  equation.
\newblock {\em Phys.\ Rev.} \textbf{D 44}, 1067 (1991).

\bibitem{Pachuckietal96} K. Pachucki, D. Leibfried, M. Weitz, A. Huber, W. Konig and T. W. Hansch. 
\newblock Theory of the energy levels and precise two-photon spectroscopy of atomic hydrogen and deuterium. 
\newblock {\em J. Phys. B: At. Mol. Opt. Phys.} \textbf{29}, 177 (1996).

\bibitem{Udemetal99} T. Udem, J. Reichert, R. Holzwarth, T. W. H\"ansch. \newblock Accurate measurement of large optical frequency differences with a mode--locked laser. 
\newblock {\em Opt. Lett.} \textbf{24}, 881 (1999).

\end{thebibliography}
\end{document}